\documentclass[acmtog, authorversion]{acmart}

\AtBeginDocument{%
  \providecommand\BibTeX{{%
    \normalfont B\kern-0.5em{\scshape i\kern-0.25em b}\kern-0.8em\TeX}}}

\copyrightyear{2023} 
\acmYear{2023} 
\setcopyright{acmlicensed}\acmConference[FAccT '23]{2023 ACM Conference on Fairness, Accountability, and Transparency}{June 12--15, 2023}{Chicago, IL, USA}
\acmBooktitle{2023 ACM Conference on Fairness, Accountability, and Transparency (FAccT '23), June 12--15, 2023, Chicago, IL, USA}
\acmPrice{15.00}
\acmDOI{10.1145/3593013.3593971}
\acmISBN{979-8-4007-0192-4/23/06}

\usepackage{graphicx}
\begin{document}

\title{Broadening AI Ethics Narratives: An Indic Art View}


\author{Ajay Divakaran (co-first author)}
\affiliation{%
\institution{SRI International }
\country{USA} }

\author{Aparna Sridhar (co-first author)}
\affiliation{%
\institution{Independent Researcher }
\country{India} }

\author{Ramya Srinivasan (co-first author) }
\affiliation{%
\institution{Fujitsu Research of America }
\country{USA} }

\renewcommand{\shortauthors}{}

\begin{abstract}
Incorporating interdisciplinary perspectives is seen as an essential step towards enhancing artificial intelligence (AI) ethics. In this regard, the field of arts is perceived to play a key role in elucidating diverse historical and cultural narratives, serving as a bridge across research communities. Most of the works that examine the interplay between the field of arts and AI ethics concern digital artworks, largely exploring the potential of computational tools in being able to surface biases in AI systems. In this paper, we investigate a complementary direction--that of uncovering the unique socio-cultural perspectives embedded in human-made art, which in turn, can be valuable in expanding the horizon of AI ethics. Through semi-structured interviews across sixteen artists, art scholars, and researchers of diverse Indian art forms like music, sculpture, painting, floor drawings, dance, etc., we explore how {\it non-Western} ethical abstractions, methods of learning, and participatory practices observed in Indian arts, one of the most ancient yet perpetual and influential art traditions, can shed light on aspects related to ethical AI systems. Through a case study concerning the Indian dance system (i.e. the {\it `Natyashastra'}), we analyze potential pathways towards enhancing ethics in AI systems. Insights from our study outline the need for (1) incorporating empathy in ethical AI algorithms, (2) integrating multimodal data formats for ethical AI system design and development, (3) viewing AI ethics as a dynamic, diverse, cumulative, and shared process rather than as a static, self-contained framework to facilitate adaptability without annihilation of values
(4) consistent life-long learning to enhance AI accountability
\end{abstract}




\maketitle

\section{Introduction}
Incorporating interdisciplinary perspectives is an essential requisite for the design of ethical artificial intelligence (AI) systems \cite{raji:pedagogy,romm:interdisciplinarity,krafft,patel}. From humanities and social sciences to law and policymaking, a growing body of excellent works have elucidated important lessons from diverse fields for ethical AI system design and evaluation. For example, works from moral and political philosophy have been leveraged to highlight emerging debates concerning fair machine learning \cite{binns}, to shed new light on human practices by situating technology within the human, social, and the political goals \cite{bietti}, and to suggest revised approaches in conceptualizing and operationalizing algorithmic recourse \cite{venkatasubramanian}. Studies have examined how document collection practices in archives can inform data collection in machine learning \cite{jo}. Principles from quantitative social sciences have been proposed as frameworks in understanding fairness in computational systems \cite{jacobs}. Lessons from feminist studies have been used to offer guidelines on how to conduct research on explanations \cite{li}, to inform data collection process in AI systems \cite{leavy}, and to study issues related to algorithmic oppression \cite{hampton}. In \cite{muller}, the authors explore the potential of participatory design fictions in learning about AI ethics. Findings from theories of parenting have been used to inform new ways of designing and developing autonomous agents \cite{croeser}. Issues concerning governance of AI technologies have been discussed leveraging design lessons from history \cite{cihon}.

In addition to the aforementioned diverse disciplines, the field of arts is regarded to play an important role in shaping fairness and accountability of AI systems \cite{beth, srinivasan, aies}. As stated in \cite{tim}, art is {\it `a form of technology that contributes to knowledge production by exemplifying aspects of the world that would otherwise go overlooked'}. It is also believed that art facilitates cultivation of `moral knowledge', i.e. knowledge about what is right and what is not right \cite{young}.  The authors in \cite{novitz} state that art also engenders empathy by creating a means of direct experience to compare different viewpoints about the world. Art is regarded as a powerful language that can bring about symbolic transposition \cite{leroi}, and is perceived to be a form of communication with the public \cite{aies}.

Scholars posit that the field of arts has the potential to transcend the limitations and distortions of dominant ethical paradigms derived from the sciences and economics by being able to showcase plurality of values, by elucidating procedures as opposed to just outcomes, and by emphasizing the role of participatory design \cite{tasioulas}. Researchers argue that artists have historically deployed new technologies in unexpected and often prescient ways \cite{shaken} and have been interpreted as vanguards: of new ideas, techniques, and cultural practices \cite{francastel,stark}. Furthermore, the authors in \cite{stark} argue that artists engaging with data analytic tools and computational techniques can share their opinions on the changing and nuanced ethical questions faced by those who use data and AI algorithms in their work. For instance, American artist Rashaad Newsome tells stories of racial injustice through the lens of colonialism to highlight existing shortcomings of AI models \cite{jensen}. In \cite{walsh}, the authors leverage AI generated art to teach middle school students about AI. Inspired by the `theory of new aesthetics' \cite{leech}, artist Avital Meshi deconstructs the notion of `whiteness' in face recognition algorithms by using performance as a tool \cite{meshi}. In {\it ImageNet Roulette}, an art project designed by researcher Kate Crawford and artist Trevor Paglen, biases associated with large scale image datasets and their adverse consequences are exposed \cite{paglen}. 

A majority of works that examine the interplay between arts and AI ethics concern digital artworks, largely exploring the potential of computational tools in supporting artists to surface issues related to bias and discrimination in AI systems. However, there is little to no work that examines the potential of {\it human-made} artworks from {\it non-Western} contexts in informing the FAccT community---Non-Western art forms can potentially bring to light several novel and unique ethical abstractions that are relevant in local geographies. Given that most studies concerning AI ethics are influenced heavily by Western cultures in their definitions, axes of discrimination, and philosophical roots \cite{nithya}, understanding non-Western socio-cultural contexts becomes even more pertinent \cite{hanna, abebe}. Through this work, we hope to address this gap by studying how art forms from India (here we consider, in a breadth first fashion, a large subset of Indian arts that is inspired by traditional works),---a non-Western country with diverse and rich cultural history, constituting nearly 17 $\%$ of the world's population, and which has the largest diaspora in the world--- can inform issues related to AI ethics. In particular, we explore the following research questions (RQs):

\begin{itemize}
    \item {\bf RQ1:} What kind of ethical abstractions can be learned from  Indian arts?
   

\item {\bf RQ2:} Can Indian arts shed light on new forms of participatory approaches that can be useful in the AI pipeline?


\end{itemize}

In order to study the aforementioned research questions, we conduct interviews of artists/practitioners, art scholars, researchers, and art aficionados. Given that information about certain forms of arts, especially performing arts like dance and music, cannot be analyzed purely based on literature reviews \cite{venugopal}, and more broadly, given that interfacing with practitioners and other domain experts can better illumine aspects related to the process of art information creation, sharing, and distribution \cite{srinivasan1}, we adopt an interview based approach. Responses from sixteen interview participants whose background span diverse Indian art forms such as Indian classical music (Carnatic and Hindustani forms), Indian style painting (silk cloth painting, Thanjavur paintings), Indian dance (Kuchipudi and Vilasini dance forms), Kolam (traditional geometric floor drawings) among others, are summarized in the context of the aforementioned RQs. 

Through a case study concerning the Indian dance system, also known as the {\it `Natyashastra}, we analyze potential pathways towards enhancing ethics in AI systems. Insights from our study outline the need for (1) incorporating empathy in ethical AI algorithms, (2) integrating multimodal data formats for ethical AI system design and development, (3) viewing AI ethics as a dynamic, diverse, cumulative, and shared process rather than as a static, self-contained framework to facilitate adaptability without annihilation of values
(4) the need for consistent life-long learning to enhance AI accountability

The rest of the paper is organized as follows. We first provide a brief review of Indian art forms outlining their history, influence, and relevance in the context of this study. We then provide details of the interview process starting from questionnaire structure and design to details pertaining to respondent selection and data consent. We then summarize the interviews responses in the light of the research questions. Next, we consider Natyashastra or the Indian dance system as a case study to recommend potential pathways towards enhancing AI ethics. Finally we close by discussing limitations and future work.

\section{Understanding Indian arts}
Scholars and historians of Indian art have long grappled with the problem of describing and understanding a vast diversity of forms of artistic expressions, over vast periods of time and space, motivated by a wide variety of psychological and philosophical traditions. Some have got around this by providing a chronology over time of the evolution and development, while others have tried to identify the chief features of particular schools or styles within each genre \cite{kapila}. 

India has a long history of art and craft \cite{dhar}. The art of the Indus Valley Civilisation (Bronze Age), which included parts of Northern India, was depicted through pottery, seals, jewellery, sculpture, etc. Most regions of India have preserved their art forms for over 3000 years through informal local communities and guilds. Commonly known as folk art, the art forms of these informal local communities are less structured and more organically imbibed than their classical traditions. Stories based on local idioms are woven into songs, paintings, puppetry, weaving, sculptures, and more. 


These traditions have sustained over time in physical form across myriad architectural, sculptural, and other art forms through canonical texts as well as through oral transmission of specific skills and knowledge over generations. India’s oldest texts - the Vedas - speak of two kinds of knowledge. The Mundaka Upanishad (an ancient Sanskrit Vedic text in the Atharva Veda) talks about {\it `apara'} knowledge (superficial knowledge, amenable to linguistic analysis) and {\it `para'} (higher knowledge, which is related to consciousness). Traditional sciences that one learns through formal training (known as {\it `siksa'}) is apara, whereas deeper knowledge that arises out of inner inquiry is para or also referred as {\it `vidya'}  \cite{rajgopalan}. The Indian arts, like much of all Indian knowledge systems, have to be approached from this perspective.

Elaborating on the continuum in Indian arts, in her essay `Inter-relationship of the Arts' \cite{vatsyayan}, Indian art historian Kapila Vatsyayan  writes, {\it ``The continuity could be maintained only because the tradition itself had an in-built paradigm of facilitating change, constantly adjusting itself to a contemporaneity of time and place while adhering to certain underlying principles which were perennial and immutable. The dynamics of the still centre and the ever kaleidoscopically changing movements of styles, schools and artistic expressions were sustained in different milieus and at different levels of society through different diverse media. The aesthetics emerged from a shared world-view with the acceptance of diversity in form." } 







Moving away from a chronological or even a style-based study of art, there has been an attempt by modern artists and scholars to look at Indian art as a manifestation of the `inner self' or the common substratum, through external symbols, motifs, and sounds. In his book `Towards Ananda - Rethinking Indian Art and Aesthetics', author Shakti Maira points out two reasons as to why it is important to integrate Indian arts into the global landscape, which would also include technical fields like AI. The first is that Indic culture has preserved art in the lives of the common man. The second point he makes is that Indian aesthetic traditions, like its Vedantic and Buddhist traditions, have dug deep into the human mind and into the nature of human experience and that {\it ``they offer a different and important vision and purpose for art and its practices"} \cite{maira}. 

The current dominant philosophy of Postmodernism postulates that all languages and art styles are cultural constructs conditioned by prevailing social circumstances and there are no pure meanings and no one reality \cite{artstory}. Author Shakti Maira counters Postmodern epistemologists arguing that it is not just the language and art styles that define experiences. He goes so far to say that {\it ``While the human experiences that Indian artists refer to are universal, yet there is an `Indianness' in the focus and stress placed on some experiences in preference to others. For example, the emphasis on inner experiences - emotional states and movement of consciousness over sensory experiences and contact with the outer world"} \cite{maira}.

The aforementioned emphasis on the inner experience is evident across works concerning Indian aesthetics written in the context of art-making. The concept of {\it `Rasa'} or aesthetic flavor was introduced by Sage Bharata (believed to be the father of Indian theatrical forms, who lived roughly between 200 BCE and 200 CE, but estimates vary between 500 BCE and 500 CE) in his seminal work the {\it `Natya Shastra'}, a Sanskrit treatise on performing arts. According to the Natyashastra, there are nine principal emotional states, human feelings, or moods, known as {\it`rasas'} - {\it shringara} (i.e. beautiful), {\it hasya} (i.e. humorous or comic), {\it karuna} (i.e. compassion, empathetic) {\it raudra} (i.e. anger), {\it vira} (i.e. brave, heroic), {\it bhayanaka} (i.e. fearful), {\it bibhatsa} (i.e. odious), {\it adbhuta} (i.e. astonishing) and {\it shanta} (i.e. peaceful). 
In the 11$^{th}$ century, Abhinavagupta, a famous theoretician on aesthetics, built a comprehensive theory of the rasas in which art and the aesthetic experiences were postulated as an outpouring of energy, creativity, and imagination of the artists and their {\it rasikas} (i.e. audience) \cite{higgins}.


The Indian way of looking inwards is brought out by renowned theatre scholar Richard Schechner in his article on Rasaesthetics \cite{schechner}, where he mentions the difference between Stanislavsky-based Euro-American acting and the Natyashastra. Schechner states that in Euro-American acting, {\it ``one does not play an emotion ; one plays the `given circumstances', the `objectives', the `through-line of action,'  the `magic if'.  If this is done right, `real' feelings will be experienced and `natural' emotions will be displayed"}. Thus, in Western theatre, both American and European, the plot and the text are primary, and it is believed that the story automatically creates the required emotions. On the contrary, across Indian theatre and all forms of performing arts, the emphasis is on emotions with the plot being secondary.
According to Schechner’s interpretation of the Natyashastra's rasic system, one can work directly on the emotions--- {\it ``mixing them according to `recipes' known to the great acting gurus (i.e., teachers) —or even by devising new recipes. From a Stanislavskian vantage, such direct work on the emotions will result in false or mechanical acting. But anyone who has seen performers thoroughly trained in the Natyashastra's rasic system knows these performers are every bit as effective as performers trained in the Stanislavsky system"} \cite{schechner}. In Schechner’s study of Aristotle’s Poetics and the Natyashastra, he says that many Indian artists subscribe to the ideal of a theatre that integrates drama, dance, and music. Traditional Indian theatrical genres accomplish this integration in ways that privilege dance, gesture, and music over the plot, thereby emphasizing the associated emotions along with other relevant cognitive and aesthetic elements \cite{schechner}.


The spiritual and the aesthetic go hand in hand in India. In the book `Rasa: Performing the Divine in India', author Susan Schwartz writes---{\it ``It is more fruitful to state that the goal of the aestheticians, from Bharata-Muni onward, has been to facilitate a transformation—of the artist, the audience, and ultimately the world—that may only be understandable from the perspective of religion. So central has the religious context been to understanding and achieving the goals of performance that it is possible to study the religions of India through her performing arts"} \cite{susan}. 


One of the greatest scholars of Indian art history Ananda Coomaraswamy (1877-1947) in his article `The Aims of Indian Art' \cite{ananda} says that there is a close analogy between the aims of art and theoretical science---{\it ``The imagination is required for both; both illustrate that natural tendency to seek the one in the many, to formulate natural laws, which is expressed in the saying that the human mind functions naturally towards unity. The aim of the trained scientific or artistic imagination is to conceive (concipio, lay hold of) invent (invenio, to light upon) or imagine (visualize) some unifying truth previously unsuspected or forgotten. ... Ideal art is thus rather a spiritual discovery than a creation. It differs from science in its concern primarily with subjective things, things as they are for us, rather than in themselves. But both art and science have the common aim of unity; of formulating natural laws."} The synthesis of Indian thought is thus understanding the one across many.

Given this broad outline, we have tried to capture the essence of an ancient artistic ethos by interviewing artists and scholars who are engaging with a wide variety of Indian artistic traditions. There is a common function among various art Indian art forms, which is to transform both the artist and the audience using the art form as the vehicle to experience such transcendence.  We posit that this functional aspect can be valuable in the context of AI ethics. The section below describes our interview methodology used to study Indian arts, covering both its form and function.

\section{Methodology}

In order to study the research questions, we conducted semi-structured interviews. The rationale behind semi-structured interviews is as follows. First, in the Indian context, the transmission of knowledge in certain forms of arts, especially the performing arts such as dance, music, and theatre, is largely oral. Thus, one can understand the pedagogical foundations by speaking to artists who have worked with generations of masters in the hallowed {\it `Guru-Shishya parampara'} (i.e. teacher-student traditions) where information is passed on from the preceptor to the pupil. Second, scholars premise that there are gaps in existing literature on documenting different aspects of development of the Indian arts \cite{venugopal}. A combination of semi-structured interviews coupled with literature review therefore helps in gathering a more comprehensive picture of the Indian art scene. 

We conducted semi-structured interviews across a variety of people engaged in different forms of Indian art. These included artists, art history scholars, AI researchers who have trained in an Indian art form, and cross-disciplinary researchers who have studied topics at the intersection of the arts and the sciences. The participant pool that we reached out to was determined so as to be representative of 1) gender, 2) age, 3) years of practice, 4) the art form, and 5) the region in India they hail from (which in turn has a bearing on training and aesthetics). We also contacted people who did not have Indian origins, but who had extensively studied an Indian art form. All the potential participants were contacted via 
email and through their social media profile. In total over thirty people were contacted, out of which we received response from sixteen. 
The experts who participated in the study include vocalists across different forms of Indian music (Carnatic classical and Hindustani), instrumentalists (Indian instruments Mridangam, Sitar, Sarod, Veena), sculptors, painters (silk cloth painting and Thanjavur painting style), dancers (Kuchipudi and Vilasini Indian dance forms), Kolam artists (a traditional form of geometric floor drawings), scholars of Indic knowledge systems, and computer science/AI researchers. Participants were distributed across continents--North America (USA), Europe (Sweden and Netherlands), and Asia (various parts of India).  

Participation in the study was voluntary. The respondents were informed of the interview's purpose; they were provided with some context and illustrations highlighting the need for fairness and accountability in AI systems. The experts who participated in the study consented to sharing the information provided by them in an academic paper or journal. Participants were asked some personal information such as their name, location, and email so that we could contact them if we needed additional clarifications from them.  

The interview was journalistic in nature - exploratory, topical, cross examining conventional premises \cite{miller,svyetlov}.  The list of interview questions was shared with the participants via a Google form prior to the interview allowing the participants to share some of their thoughts through the form before the interview. Participants were also asked to indicate their engagement with arts--as art scholars, art researchers, practitioners/artists, aficionados, etc. The questions spanned areas that can described as art aesthetics, art education, art history, art psychology, and artificial intelligence. Although the questions were primarily motivated by the research questions, they were designed to be personal, introspective, and open-ended, thereby allowing flexibility in interpretation and adaptation. There is an element of subjectivity to the method as close-ended research cannot capture the highly subjective, creative world of artists. While the pedagogy of the Indic artistic tradition is extremely principled and replicable, the artist adds their own {\it `manodharma} (i.e. improvisation) which makes every artistic work unique. 

First, leveraging our backgrounds across fields like Indian art history, music, social science, and artificial intelligence, we designed a set of questions to understand how the learning, training, execution, and reception of an Indian art form, which happens over a long period of time, can be used to address issues related to AI ethics. We further refined the questions and added new questions after consulting scholars of Indian arts and artificial intelligence. We provide a brief description of the question topics below.

In order to understand how people across different disciplines perceive artificial intelligence, we included a question that asked the respondents to explain what they think AI is and list the domains that have influenced their understanding of AI. Some artists and art scholars whom we contacted but who did not participate in the interview cited that they were not too familiar with AI related aspects to participate; thus including this question was mutually beneficial- it helped the experts to determine their participation, and helped us in obtaining informed responses. 

We included specific questions pertaining to the art traditions and learning to understand if the process of traditional art information/knowledge sharing can inform the AI community of new forms of learning and representations. We also included questions related to collaborations between artists, how artists connect with the audience, how they are able to achieve harmony in renditions, and how they adapt to changing times to understand if such artistic practices can shed light on new forms of participatory approaches for AI systems. In order to understand if Indian arts can inform the AI community of new ethical concepts, especially those that are relevant in non-Western contexts, we also asked the respondents to describe unique ethical concepts seen in Indian arts. Finally, we also asked if the respondents had any message for AI researchers. The most common response to the interview questionnaire was that some questions were rather deep, requiring long response times and thought. Interviews were primarily conducted over audio calls and the details of the interview calls were manually transcribed to identify the main themes in the responses. The study was approved by the institutions of the authors. 

\section{Summary of Interview Responses}
In this section, we provide a summary of the interview responses in the light of the RQs considered for the study, along with our brief extrapolation of the viewpoints suggested in the responses. 
\subsection{Discussion of RQ1} 
{\it What kind of ethical abstractions can be learned from the Indian arts?}

In order to examine this RQ, in our interview questionnaire, we had included a question wherein the respondents were asked to describe the process of information gathering, learning, and sharing in their art forms, and what were the distinct values associated with their art form and learning process (when compared to other art forms and learning practices). We also asked them to describe the checks and balances inbuilt in the training and delivery of their art forms, and to describe how they maintain consistency with traditional guidelines and practices. We also had included a question asking the participants to describe, if any, ethical abstractions or concepts that they think is unique to Indian arts. We had also included an open ended question that asked the participants to share their opinions/recommendations with respect to enhancing ethics in AI systems. Below, we summarize these responses.

In response to the question concerning unique ethical abstractions or concepts from Indian arts, one participant mentioned that the spirit of seeing the universe as one family or the principle of {\it `Vishwa Kutumba'} that is exemplified in Indian philosophy and arts could be valuable. Another respondent mentioned that the principle of {\it `Paropakaraya punyaya payapa para peedanam'}, which roughly translates to `ensuring welfare for all and penalizing immoral acts' could be valuable for the design of ethical AI systems.

Yet another respondent mentioned that unlike Western philosophies which are largely driven by rule based frameworks and which are based on the premise that individuals are different, Indian philosophy recognizes the commonalities across people, and that this ideology is reflected in the Indian arts as well. According to this participant, ethical is anything that evokes the inner reality within an individual whereas unethical is that which revolts this process. The participant further added that the Indian art depicts the common thread across all life forms. This spirit has been extensively reflected across works of many Indian poets and artists who emphasized that the physical basis of all life forms is the same, independent of factors such as gender, religion, caste, and so on \cite{narayan}. 

Echoing the aforementioned thought, one participant pointed out that Indian art goes beyond binaries and cited the example of {\it `Ardhanarishvara'} (a Sanskrit word meaning 'Lord who is half Woman') which is a composite male-female figure in the Indian sacred pantheon. The participant added that across many Indian and Southeast Asian sculptures, icons are decorated combining the male and female aspects, which are perceived to be inseparable. 
In a similar vein, contrary to Western notions of race wherein fair complexion (white) is considered to belong to an advantaged group, poets of ancient India considered dark complexion as something vibrant and throbbing with life and exuberance. For example, almost all of the female characters in Kalidasa’s (considered to be one of India's greatest playwrights and dramatists) masterpieces were dark complexioned \cite{sumit}.

In response to the question about what is distinctive in their art style and art philosophy, another participant said that there was a deep connect between the meaning of art and everyday life - which is manifested in performative traditions even today. The participant added that Indian art has avoided the pitfall of art becoming elitist or merely an intellectual exercise by connecting it to the daily lives of people. 
The Indian arts offer a multi-dimensional image of the human mind. Studying the basic guidelines and methodologies of the making of Indian art, said one participant, might be valuable in incorporating its core ideas to other disciplines, including AI. Giving the example of Kolams which are sacred geometrical drawings, another participant outlined how women across India clear the front of their houses and draw Kolams ---geometrical pattern of dots and lines--- using rice or rock powder to energize the house. It is an ancient artistic ritual which while being temporal (as it can be easily erased) has also remained in the collective memory across generations of Indians. The dots are said to represent the challenges of life and people have to weave their way out of the maze, in a harmonious manner as depicted by the synchronous Kolams. The aim is to look for beauty, context, content, and core in all aspects of everyday life. 

Thus, understanding the multi-dimensional nature of intelligence and incorporating cognitive aspects like empathy and emotions can be beneficial in the context of training ethical AI systems. Objective rules are useful in as much as they can serve as a baseline, but subjective elements like empathy and emotions have to be considered in order to enforce a sense of purpose, stability, and well being of all stakeholders involved in the AI pipeline.  Understanding how humans traverse through everyday decision making (such as through past experiences, ancient knowledge repositories, formal education, etc.) can help enhance accountability in AI systems. 
Further, the aforementioned responses emphasize the need to look beyond rule based abstractions for characterizing ethics in AI systems. In fact, ancient Indian texts advocate several factors for the observance of justice. These include aspects such as compassion, forgiveness, patience, truthfulness, absence of anger, piety, sanctity, non-violence, and control of senses. These aspects have been illustrated across different art forms. As also pointed out in \cite{nithya}, notions beyond fairness such as restorative justice \cite{boyes,inayatullah} and those based on moral foundations such as purity and compassion \cite{jesse} can help shed light on philosophical viewpoints that are relevant in local cultural and geographical contexts.

\subsection{Discussion of RQ2}
{\it Can the Indian arts shed light on new forms of participatory approaches that can be useful in the AI pipeline?}

In order to analyze this RQ, in our interview, we had included questions concerning how artists collaborate with co-artists and accommodate their views to achieve harmony in their renditions, how they are able to reconcile contradicting opinions with their co-artists, how they cater to the requirements of diverse audience, how they adapt to changing times, and the extent of freedom they have in interpreting the art works of other artists. Below, we summarize the responses to these questions.  

In response to the question concerning reconciliation of contradicting opinions (which often becomes necessary in settings involving stakeholders with differing requirements such as in an AI pipeline), one participant cited the example of the proliferation of Harihara sculptures from the 5$^{th}$ century onwards and mentioned that often art was seen as a vehicle to transcend differences. The participant mentioned that the Harihara sculptures are indicative of the forces of fusion and syncretism in the Indian society.  Ontologically, Harihara signifies that there is no metaphysical difference between the two principal divinities of Hinduism -Vishnu (Hari) and Shiva (Hara). Epistemologically, it represents the idea that diversity emerges from unity before merging itself into unity again. Another example of syncretism given by the participant concerned a medieval sculpture at the Indian Museum in Kolkata which represents Harihara flanked by Buddha and Surya. In this, a single sculpture combines the four most important philosophies – Saiva, Bhagvata, Saura, and Saugata.  

Adding to the aforementioned line of thought, another participant quoted the Sanskrit phrase {\it `spardhaaya vardhate vidya'}, which roughly means competition enhances knowledge. Elaborating on this point, the participant said that understanding the benefits of complementary viewpoints helps in  convergence of divergent art forms, and in turn, helps in presenting a collaborative performance as one aesthetic entity. 

In response to a question concerning how knowing about other art forms helps an artist to be more creative and accommodate other art forms, one participant said that being aware of multiple art forms and immersing oneself in various art forms to the extent possible, enhances accommodation. A few other participants mentioned that various art forms are intrinsically interconnected and an awareness of the same can help them in engaging creatively with their co-artists. Elaborating this point by taking the example of music based art forms, one participant mentioned that Indian music reflects the metaphysical idea at the heart of Indian philosophy which is that humans are all fragmented parts of a whole, ultimately united as one. According to the participant, this makes one approach music as a means to understand this oneness rather than merely as individual expressions of creativity. The participant added that this attempt at connectivity can be seen in the {\it Raga/tala} system of Indian music, which is connected to nature and the environment, and ultimately to the essence of the universe. The participant gave the example of {\it `svaras'} in Indian music which refer to the musical notes on an octave, which are believed to have origins in the music of animals (notes: Re, Ga, Dha, Ni) and birds (notes Sa, Ma and Pa).

In response to a question concerning the extent of freedom artists have in interpreting works of other artists and incorporating the same in their renditions, one participant mentioned that respect to the values of the original works is imperative. Another participant said interpreting an art piece necessitates an analysis of the literal meaning and context of the piece, the cultural, philosophical and ideological leanings of the artist, the socio-political context, the layers of meaning in the art composition like the {\it `Padartha'} or the literal meaning, {\it `Vakyartha'} or the grammatical embeddings, {\it `Gudhartha'} or the hidden meaning, {\it `Visheshartha'} or the nuanced higher meaning, the symbolism, imagery, and the sub texts. The participant mentioned that all these put together helps in understanding the totality of the art piece. 

In response to a question that asked how artists adapt to changing times, one  participant quoted one of India's oldest poets Kalidasa who says  {\it `puranamityeva na sadhu sarvam, na chaapi kavyam navamitya vadhyam'}, a Sanskrit verse which means not everything ancient is always good and not everything modern is bad. The participant mentioned that artists have twin responsibilities - to cater to the changing aesthetic sensibilities of the connoisseurs and secondly to mould the aesthetics of the audience and raise their standards of understanding the underlying values and beauty. This twin purpose should probably serve AI systems well too.

All artists have to adapt to changing times and even react to change. One participant observed that the arts, particularly performing arts, cannot or should not freeze in time.  The tradition becomes the unchanging core which gives the artist the capability to face changes. The participant said that while modern tools and technologies can be helpful in furthering traditions and traditional art forms, the disruption can be caused only by the user, not the times or the changes it brings to the situation.

Another participant said the way in which art can both embrace and shock diverse audiences is an important metaphor for how AI could be designed. The participant said a lot of AI technologies today facilitate the creation of bubbles - echo chambers of niche interests and ideologies, and added that AI must move away from identifying and grouping similar subsets, but should open new ways at looking at the world. They said that an AI system should act like a friend in school who was a tastemaker, who was not interested in recommending music that the respondent liked, but instead would deliberately throw in something that one had never heard before and thereby open one's mind to new experiences.  

An oft quoted metaphor is that of the wheel, where the core at the center is fixed but the rim moves on. Most of the participants subscribed to this notion of the continuity of the artistic tradition through changing times.
This characteristic of continuous adaptability without annihilation of diverse values is also needed in AI systems. Adaptability is thus a cumulative process that synchronously blends diverse principles in a gradual manner so as to accommodate change while maintaining continuity, which can be valuable in enhancing robustness of AI systems. 

Based on the above responses, it can be inferred that recognizing and respecting subjective viewpoints is an essential requisite in any participatory framework. It is important to note that subjective viewpoints are inherently dynamic: they vary from person to person, place to place, and time to time. Thus, incorporation of subjective viewpoints can enhance the applicability of AI systems. Further, understanding the broader socio-cultural settings of a problem, understanding domain-specific requirements and challenges, and striving for harmony even in presence of differing stakeholder opinions can therefore be useful participatory paradigms to consider in the AI pipeline.

\section{Case Study: Natyashastra}
In this section, we will mostly consider Natyashastra as a running illustration to put forth the unique aspects that the Indian arts can convey in the context of ethical AI system design and development. As discussed in Section 2, Natyashastra is a seminal and ancient Indian text that is regarded as a comprehensive practical guide for a broad range of art forms such as music, dance, theatre, sculpture, and drama. Furthermore, it is said that Natyashastra laid down the essentials of performing arts as a representation of the ways of the world, the nature and attitudes of the people, their ways of behavior and manners of speech, providing necessary moral guidelines for actors in rendering flawless performances.  It is said that the Natyashastra shows how the audience can feel for a moment the transcedental experience which can form the basis of all human aspiration \cite{ghosh}. For these reasons, we primarily consider Natyashastra as for our analysis. Table 1 summarizes the findings from the analysis by providing a mapping of the Natyshastra principle (leftmost column) to the broad ethical AI design idea (center column) and potential pathways in which the idea can be explored/developed (rightmost column).

\subsection{The need for empathy in characterizing AI ethics} 
Indian art and philosophy offers different interpretations of intelligence and ethics, which have over the last few decades gained acceptance in modern psychology \cite{singh}. In the Indian view, intelligence refers  to `waking  up,  noticing,  recognizing, understanding, comprehending, and caring. {\it Buddhi} or intellect includes such things as determination, mental effort, feelings, and opinions in addition to intellectual processes such as knowledge, discrimination, and decision making \cite{singh}. As noted by one of our  interview participants, Indian dance and music not only demonstrates the kind of intelligence seen in computational systems--- such as higher calculative ability (in Indian rhythm), superlative memory (evidence includes learning of thousands of songs through oral transmission), decision making capacity (while innovating on the spot), high speeds of action (in responding to co-artists), using the prosodic structure well--- but it also demonstrates other capabilities often ignored by computational systems. These forms of intelligence have been described as {\it prathibha} (keen intellect), {\it prajna} (transcendental wisdom), {\it Vak} (speech), and {\it rasas} (emotions).

Emotions or the rasas (discussed in Section 2) form a primary premise of Indian dance system \cite{ghosh}. One participant who is a dancer mentioned that there is an established methodology to understand, codify, and express emotions in the Indian dance forms through a five-tier structure of emotions which includes {\it `Vibhava'} or the determinant, {\it `Anubhava'} or the consequent, {\it `Sthayi Bhava'} or the dominant emotion, {\it `Vyabhichari Bhava'} or the transient emotions, and {\it `Sattvika Bhava'} or natural bodily responses to emotion. Thus, rasas offer a framework to understand causes and determinants of events as well as help in analyzing their transitions and effects. Furthermore, Natyashastra also describes that supporting causes (of usually the hero or the heroine or such objects) are of two types {\it `vishaylamban'} (person or object of the rise of an emotion or the person or object for whom the emotion is awakened) and {\it `ashramban'} (person in whom the emotion is awakened). This kind of fine-grained analysis is beneficial for the design of ethical AI systems in that they can help in analyzing the benefit of the system for various stakeholders--what purpose it is serving, for whom, for what, etc.  Rasas also facilitate communication between the actors and the audience, serving as a pathway to help both the actors and the audience empathize with each other's opinions. Natyashastra prescribes rasas as the essential guidelines that actors should diligently follow in order to creditably acquit themselves in their roles \cite{ghosh}. We posit that infusion of rasas or more broadly empathy in AI systems will allow perspective taking, feeling, action, meaning, understanding, imagination, socialization, and care-taking. These aspects in turn will result in AI systems that are aligned with human values, thereby facilitating the creation of trust-worthy systems.

One possible way in which empathy can be infused towards enhancing ethical values in AI systems is by means of intervention aided reinforcement learning. By this we mean human-guided interventions in a reinforcement learning framework to optimize for aspects of empathy such as care, understanding, social welfare, and feelings. Research in robotics can provide useful cues in this regard \cite{wang2018}. Causal reinforcement learning frameworks could also be leveraged to learn when to intervene \cite{zhang2020}. Alternatively, the problem could be formulated as a restless multi-armed bandit problem such as in \cite{biswas2021}, where each stakeholder in the AI pipeline is assumed to transition from one cognitive state to another depending on the intervention, and the goal is to find the transition probabilities optimizing for empathy (which is captured in the cognitive states). 

\subsection{Integrating multimodal data for characterizing ethics}
Natya is seen as a comprehensive art, fusing several disciplines including music, literature, sculpture, painting, even architecture, and {\it chikitsa shastra} (i.e. study of health). Beyond rasas or emotions, Natyashastra prescribes a variety of multimodal cues for expression. These include {\it`vacika abhinaya'} (i.e. linguisitc expression to express tone, diction, and pitch of a particular character), {\it `angica abhinaya'} (i.e. bodily gestures), {\it `aharya abhinaya'} (i.e. costumes), {\it`sattvika abhinaya} (i.e. voluntary gestures expressed by presence of tears, horripilation, etc. to express deepest emotions). It is also characterized by different modes of production such as those based on dominance of spoken words, dominance of elevated and heroic feelings, and dominance of violent and conflictual actions. Further, there are also different kinds of musical instruments used on the dance stage and songs are included at different junctures of the performance to convey different meanings (e.g., songs sung as interventions, as entertainment, and to sooth emotions of the audience). 

We posit that in order to develop ethical AI systems, a similar multimodal approach is  necessary. This not only entails providing multimodal explanations underlying an AI decision and training systems through demonstrations, but also leveraging multimodal data sources (text, image, video, audio) encompassing diverse ethical perspectives in defining, designing, and developing ethical AI systems. Existing works only address parts of the proposed strategy such as through multimodal explanations \cite{park,kanehira}, exploration of fairness related issues in multimodal settings \cite{chen,wang}, inference of norms from stories \cite{frazier}, or by using principles of imitation learning towards value alignment \cite{taylor} ; these works neither leverage multimodal data sources nor incorporate diverse ethical perspectives in their design. 
For example, there is little to no work that investigates the potential of audio signals in the context of AI ethics. As can be understood by the analysis of Natyashastra \cite{ghosh}, music and speech have the potential to characterize cause and effect, channelize events, and transform the minds of the audience. Indeed, studies in psychology have emphasized the role of sound in problems related to causal attributions \cite{gerstenberg}. These aspects are very pertinent in questions related to AI accountability. Thus, it becomes necessary to go beyond the current emphasis on a selected subset of data modalities in defining ethical notions (such as fairness based on statistical parity or equalized odds using categorical data) and leverage diverse data modalities such as audio and images as well in such characterizations whenever possible. More broadly, we believe that leveraging multimodal data during all stages of the AI pipeline---from problem formulation and data curation to algorithm design and evaluation--can be beneficial in capturing and reflecting diverse ethical values, facilitating a holistic world view.

\subsection{Facilitating Adaptation without Annihilation }

Indian art is the result of constant dialog between the creator and the audience, and therefore has evolved as the tastes of the audience have evolved. At the same time, it has preserved its essential nature, so it is still recognizable as Indian art. In other words, it is a work in progress that maintains a constant identity while also constantly changing. Part of such fluidity also stems from the integral role of improvisation in Indian art. No two Indian dance performances even by the same artist are the same since the artist improvises while adhering to overarching principles to connect with the audience. For example, in the Natyashastra, it is described that when the determinant, consequent, dominant and transient emotions all abandon their local, individual, or temporal association or limitations there is `sadharanikarana' or generalization \cite{ghosh}. Giving the example of hero and heroine in a dance play, the Natyashastra says that when such a generalization is reached, the actors are not confined within particular intervals of time and space, instead they appear before us as ordinary lover and beloved. Accordingly their love becomes the love of ordinary man and woman. It is further emphasized that only after this generalization has taken place in the mind of the audience, the rasa or the emotion can be tasted. It is said that this generalization process occurs through a unification of the primary meaning, the underlying emotion, and the resulting experience in a seemingly imperceptible manner, cumulatively \cite{ghosh}. 

Ethics offers a similar challenge of maintaining a recognizable core while adapting to changing times and unanticipated needs. AI ethics is a shared, dynamic, cumulative process as opposed to just being a static framework. AI systems need to go beyond static training sets to dynamic learning in which the model is able to adapt to the changes in the data over time while maintaining adherence to ethical principles. The generalization process described in the Natyashastra can serve as a starting point for the design of AI systems that maintain adherence to essential principles while also adapting to change. As suggested in the Natyashastras, this necessitates the unification of the primary purpose (i.e. problem under consideration), the emotion ( the stakeholders' interests), and the resulting experience (i.e. the end product and its influence on the consumers). In order to achieve this, transcending limitations of time, space, and other such boundaries is required---it therefore becomes imperative to understand the commonalities across geographies. Explicit ethical principles are also grounded in experience and can be thought of as resulting from human distillation of experience. Harmonization of ethics stemming from immediate experience and ethics stemming from years or decades of experience requires careful understanding of what is in common and what is not between the two. 

The bottom up nature of data driven learning in fact provides a bridge between explicit ethical principles and principles stemming from practice. While data-driven techniques operate bottom up, recently neuro-symbolic techniques \cite{Rocktaschel}, \cite{SikkaDASL} have been developed that impose top-down constraints on AI models. These methods use first-order logic to regularize the bottom up learning so that it conforms to background knowledge such as `a human can ride a horse', and `a horse cannot ride a human'. Similarly causal models \cite{pearl} have also been proposed which also operate on the principle of using common sense or background knowledge to guide the bottom up learning. Such techniques offer the promise of incorporating ethics top-down into the AI pipeline, since ethics is also a form of background knowledge. The advantage of neuro-symbolic techniques is that they retain the accuracy of the deep learning frameworks while injecting background knowledge. If ethics can be expressed in term of first order logic or simple causal frameworks that can capture the commonalities across geographies in a holistic manner, then very likely such methods will enable organic incorporation of ethics into AI. The challenge is of course that not all ethical frameworks can be captured in first order logic. This challenge becomes especially acute in the case of Indian art in which underlying principles are either stated at a high semantic level that are beyond current knowledge representations or are simply understood in an implicit fashion from multiple instantiations.  The challenge for AI systems going forward to is to incorporate such a holistic view of ethics into the learning framework so that ethical considerations are baked into the system. 

\subsection{Consistent life-long learning}

An important aspect in the Indian dance system is the {\it `guru-shishya parampara'} or the teacher-taught tradition, wherein a student learns the artform from the teacher over many years, sometimes even spanning a lifetime. Unlike modern day schools which are for a fixed time during the day, students in ancient India would often stay for years in {\it `Gurukuls'} for mastering various skills and Vedic knowledge. The guru-shishya tradition was a one-to-one altruistic and intimate relation in which the teacher would work individually with each student, to ensure seamless transmission of highest and the most authentic knowledge to the disciple. The teacher would aim to empower the disciple so that they would become teachers themselves down the line. On their part, the disciple had to prove that they are worthy of receiving such knowledge by demonstrating purity, fidelity, and clarity of purpose. This teacher-taught traditional Indian system was instrumental in ensuring that highest quality of teaching sustained. 

Given their similarity to teacher-taught tradition, we premise that knowledge distillation techniques and student-teacher learning can be useful analogies in the design of ethical AI systems \cite{yoon}. If one can think of the entire body of knowledge concerning ethics in the world as a large teacher model, then the goal is to incorporate the values of this large teacher model into a small student model, which in this context is an ethical AI system, as accurately as possible. Of course, this computational abstraction has limitations--it cannot possibly capture all the nuances associated with the complex body of world knowledge pertaining to ethics-- but nevertheless can be a starting point given its semblance to the teacher-taught techniques of art education. It is important to notice that the learning process in traditional art education spans many years, sometimes even decades. Thus, in order to be able to effectively leverage the benefits of such teacher-taught systems, AI models need to be trained over long regimes, possibly even requiring life-long learning \cite{parisi,lange}. A recent study supports this claim by demonstrating that ``patience" (training over large epochs) and ``consistency" (teacher and student should process the exact same data and match on a large number of support points to generalize well) are important in improving the effectiveness of knowledge distillation techniques in practical settings \cite{beyer}. Training over long regimes and ensuring consistency between the ethical models/theories out in the world and their computational abstractions will thereby help in enhancing AI ethics.

\begin{table*}
\begin{tabular}{|l|l|l|}
\hline
{\bf Principle}& {\bf Pointer} &  {\bf Potential Pathway} \\
\hline
Rasa aesthetics & Infusion of empathy  & intervention aided \\
(please see Sec 5.1 and 2 for details) && reinforcement learning\\ \hline
Abhinaya forms  & integration of multimodal data  & exploring the potential of audio-visual\\ & &  signals in understanding causal  \\ 
(please see Sec 5.2 for details) && attributions and AI accountability  \\ \hline
Sadharanikarana  & adaptation without annihilation & neuro-symbolic techniques, \\ (please refer to Sec 5.3 for details)&& causal models \\ \hline
Guru-shishya parampara  & consistent life-long learning  &  knowledge distillation techniques,  \\ 
(please refer to Sec 5.4 for details) && student-teacher models \\ \hline

\end{tabular}
\caption{{\small Takeaways from Natyashastra for ethical AI design and development: A mapping of the Natyshastra principle (leftmost column) to the broad ethical AI design idea (center column) and potential pathways in which the idea can be explored/developed (rightmost column).}}\label{tab1}
\end{table*}

\section{Limitations}
The study reflects the background of the authors and the interview participants. The study was carried out with an intention of identifying how the overarching principles of Indian arts can inform on matters related to AI ethics. We thus do not delve deep into any given Indian art form such as music or dance, and do not necessarily provide a comprehensive coverage of either Indian arts or its possible influence on AI ethics.  A wider study of Indian art might reveal aspects we have left uncovered. The technical aspects addressed here are thus only a subset of the larger requirements of ethical AI systems. An analysis of specific works of Indian art could potentially offer a rich avenue for further research. The study focused on one non-Western art form, inclusion of other non-Western art forms (beyond Indian arts) can shed light on other valuable abstractions for ethical AI system design and development. 

\section{Conclusions}
Exclusion of non-Western perspectives in ethical AI system design and development creates an imbalance in that it not only sidelines the values and principles of these communities, but also fails to understand the broader historical, cultural, and social narratives relevant to a problem, thereby resulting in adverse consequences. In this regard, the field of arts has the potential to shed light on diverse and unique socio-cultural-historical narratives, thereby serving as a bridge across research communities. In this work, through qualitative interviews of artists, art scholars, and researchers of diverse Indian art forms such as music, dance, sculpture, painting, floor drawings, etc., we analyzed how non-Western ethical abstractions, methods of learning, and participatory approaches observed in Indian arts, one of the most ancient yet perpetual and influential art traditions, can inform on matters related to AI ethics. Insights from our study outline the need for (1) incorporating empathy in ethical AI algorithms, (2) integrating multimodal data formats for ethical AI system design and development, (3) viewing AI ethics as a dynamic, diverse, cumulative, and shared process rather than as a static, self-contained framework to facilitate adaptability without annihilation of values,
and (4) consistent life long learning to enhance AI accountability.

\begin{acks}
We thank the anonymous experts who participated in the interview study for sharing their valuable insights and opinions. 
\end{acks}

\bibliographystyle{ACM-Reference-Format}
\bibliography{sample-base}

\appendix

\end{document}